# Meta-fences: blocking waves utilizing a layer of identical tiny scatters

Yunhao Zhang[1], Zhendong Sha[1], Guangyuan Su[1], Hao Zhou[1,2], Jiangzhong Yang[2], Peng Jiang[1], Yongquan Liu[1]*, Bing Li[3], and Tiejun Wang[1]*

[1]*State Key Laboratory for Strength and Vibration of Mechanical Structures, School of Aerospace Engineering, Xi'an Jiaotong University, Xi'an 710049, China*

[2]*Beijing Key Laboratory of Intelligent Space Robotic Systems Technology and Applications, Beijing Institute of Spacecraft System Engineering, China Academy of Space Technology, Beijing 100094, China*

[3]*School of Aeronautics, Northwestern Polytechnical University, Xi'an, Shaanxi, 710072, China*

*Corresponding authors, E-mail addresses: liuy2018@xjtu.edu.cn (Y. Liu), wangtj@mail.xjtu.edu.cn (T.J. Wang)

**Abstract**

Wave steering by artificial materials (for example, phononic crystals and acoustic metamaterials) is a fascinating frontier in modern physics and engineering, but suffers from bulky sizes and intractable challenges in fabrication. Here, a sparse layer of identical tiny scatters, which we call meta-fences, is presented with a non-destructive way to omnidirectionally block flexural waves in plates. The underlying mechanism is that the restraining force and moment of the scatter are tuned simultaneously to counter-balance the incident wave. Both our experimental results and numerical analysis have demonstrated that broadband wave sources ranging from 3 to 7 kHz can be segregated from the protected area by the meta-fence. In addition, the meta-fence is further assembled into a waveguide routing with an arbitrary configuration. Compared with previous isolators and waveguides, our meta-fences exhibit absolute advantages in compact size, flexible configuration, and high structural strength. The current scenario sheds light on the design of lightweight-and-strong architectures for vibration control and energy harvesting with a high efficiency, and can be extended to microfluidics, acoustics, seismology and other fields.

**Introduction**

Steering the flow of electrons, photons, and various types of waves or energy in a desired manner is a widely concerned inverse problem in modern physics and engineering. Since 1990s, phononic crystals[1,2], artificial materials composed of periodic arrangement of scatters, have been proposed to inhibit the propagation of mechanical waves based on the existence of band gaps. Then, the technique of metamaterials was introduced to engineer wavefronts by the perspective of unprecedented negative constitutive parameters (such as the bulk modulus, density, and chirality)[3-6], making it an active topic in various areas of condensed matter physics. By underpinning well-designed defects [7-9] or topologically protected features[10-15], one can guide and trap waves in a more flexible way and with robust routes. From a fundamental perspective, however, these artificial materials are applied in the form of wave propagation media, where the working wavelengths should be much smaller than the sizes of structures. As a consequence, multiple stacks of delicate microstructures are typically required to design these bulky-size artificial materials (see Fig. 1a), which may lead to intractable challenges in fabrication and practical applications.

As a more efficient and compact paradigm, metasurface[16-21] was proposed to reshape wavefronts on certain surfaces/interfaces by a single-layer assembly of subwavelength units. Based on artificial phase shifts, previous studies have already demonstrated the feasibility to transform[22-25], isolate[26,27], and absorb[28,29] elastic waves with metasurfaces. It is however noteworthy that most of the existing metasurfaces require densely arranged subunits to discretize a continuously varying phase profile, as shown in Fig. 1b. Each subunit has to be well-designed with a gradually varied configuration to fulfil the exquisite, gradient phase-shift. One needs to tailor the metasurfaces point by point through different subunits, which distinctly impede the design efficiency, availability and robustness, also requiring a combination of high-resolution fabrication techniques. To solve this problem, the concept of metagrating, an artificial surface/interface periodically arranged by only a single kind of units, was proposed in 2017 to efficiently steer wavefronts[30]. Without the need of multiple subwavelength elements to shape gradient phase-shift in metasurfaces, metagratings greatly alleviate fabrication limitations and stimulate the development of integrated optical devices[31,32].

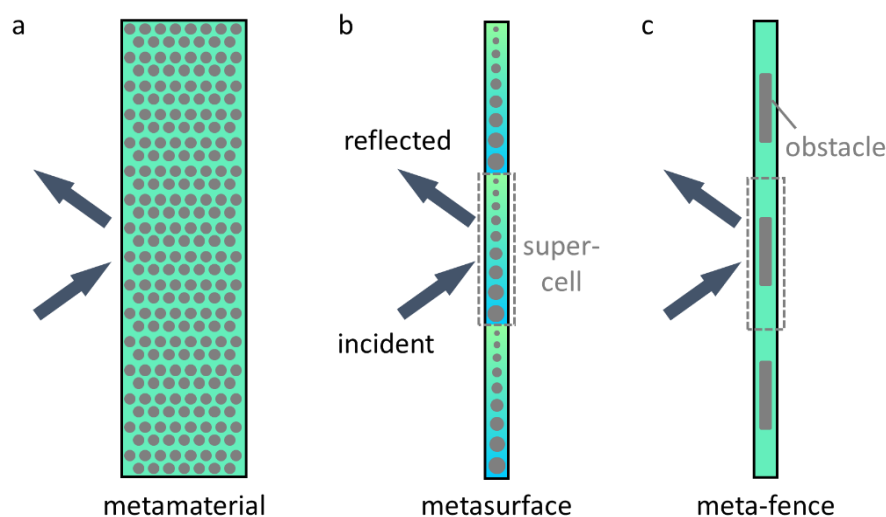

**Fig. 1 Wave-blocking based on different techniques. a** Waves can be totally reflected by the metamaterial with a band gap. The size of metamaterial should be much larger than the wavelength. **b** A structured layer is needed to stop waves based on the metasurface approach, but densely arranged subunits are required to construct a super-cell. **c** In the proposed meta-fence, only a layer of identical tiny scatters/obstacles are sparsely arranged to inhibit the propagation of waves.

In contrast to metagratings for redirecting waves, a natural question is can we design an artificial interface with uniform tiny units to completely stop the propagation of waves? To this end, mismatched (extremely large or extremely small) mechanical impedance is essentially required. For extremely large mechanical impedance, huge obstacles are competent solutions although they cannot be used in lightweight structures. On the other hand, for extremely small mechanical impedance, soft materials or structures with large porosity are used to dramatically reduce the stiffness and strength of the whole structure. A compact and nondestructive way to isolate waves and vibrations is still a long-standing challenge. In this work, we proposed a new strategy, which we call meta-fence, to omnidirectionally block waves [see Fig. 1c]. Considering the extensive applications in vibration mitigation and structural health monitoring in aircrafts, automobiles and venue buildings[33,34], we target to block flexural waves in plate-like structures in this work. Requiring neither bulky materials nor densely arranged subunits, our meta-fences work well within a broad frequency range by using a sparse array of identical tiny scatters glued on the plate. The meta-fence can further be assembled as an enclosed area to separate the wave source from the protected area. In addition, the meta-fence is also proposed to guide flexural waves along an arbitrary route. The present design enriches the design toolbox for vibration control and energy harvesting, and may shed light on designing compact wave-blocking devices in microfluidics, acoustics, seismology, etc.

**Results**

**Preventing wave propagation using meta-fences**

We first focus on a simplified 2D configuration of meta-fences to totally stop wave propagation. As shown in Fig. 2a, we consider a rectangular obstacle with height $L_h$ and width $L_w$ on a thin plate with thickness $d$. When the flexural wave normally impinges on the obstacle, additional force and moment will be generated to restrain the propagation of incident waves[35]. The restraining force and moment stimulate the bending and twisting motions of the obstacle region, respectively. Thus, the obstacle acts as a scatter to reflect and transmit the flexural waves. After detailed derivations (Supplementary Note 1), the transmission coefficient $t$ can be obtained as

$$t = \frac{i(1 + \alpha - \beta)}{(i + \alpha + \alpha i)(1 - \beta - \beta i)} \quad (1)$$

where $\alpha = \frac{\rho_{obs} A \omega^2}{4Dk^3}$ and $\beta = \frac{\rho_{obs} J \omega^2}{4Dk}$ are relative flexural and moment dynamic stiffnesses as resistances against bending and twisting motions, respectively, with $k$ the wave number, $\omega$ the angular frequency, $D$ the bending stiffness of the plate, $\rho_{obs}$ the density, $A$ the section area and $J$ the polar moment of the obstacle. The transmission $|t|$ with respect to $\alpha$ and $\beta$ is plotted in Fig. 2b. It keeps high ($|t| > 0.8$) when $\beta$ is smaller than 1, which infers that a concentrated obstacle with neglected polar moment $J$ is insufficient to hinder flexural waves. On the contrary, a low transmission ($|t| < 0.1$ for example) can be obtained when both $\alpha$ and $\beta$ are large enough ($\alpha > 3$ and $\beta > 4$) via extremely large obstacle. This condition is analogous to the low transmission in electromagnetism and acoustics in the case of extremely mismatched impedances[19]. Interestingly, we find that the flexural wave is totally reflected by the obstacle ($|t| = 0$) when $1 + \alpha - \beta = 0$ (the white line). This key finding means that a small obstacle is also possible to efficiently block flexural waves if one tunes $\alpha$ (corresponding to the restraining force) and $\beta$ (corresponding to the restraining moment) simultaneously to offset the incident wave.

To illustrate the transmission feature of flexural waves, we keep the width $L_w = 3.0$ mm of the rectangular obstacle and change its height $L_h$. As shown in Fig. 2b, both $\alpha$ and $\beta$ increase with $L_h$, but the only intersection of the black and white curves infers an optimal height of total reflection. Beyond this intersecting point, a larger barrier may unexpectedly lead to a worse isolation performance. Figure 2c quantitively shows that the intersecting point is located at $L_h = 13.1$ mm in our example (see Supplementary Note 1 for detailed material and geometric settings and the solution process). The numerical results (blue points) are in excellent agreement with our theoretical prediction, indicating nearly perfect reflection at $L_h = 14.0$ mm, where the rectangular obstacle works as a "wall" to prevent the propagation of waves. It is noted that the obstacle size is quite small compared with the wavelength ($\lambda = 39.76$ mm).

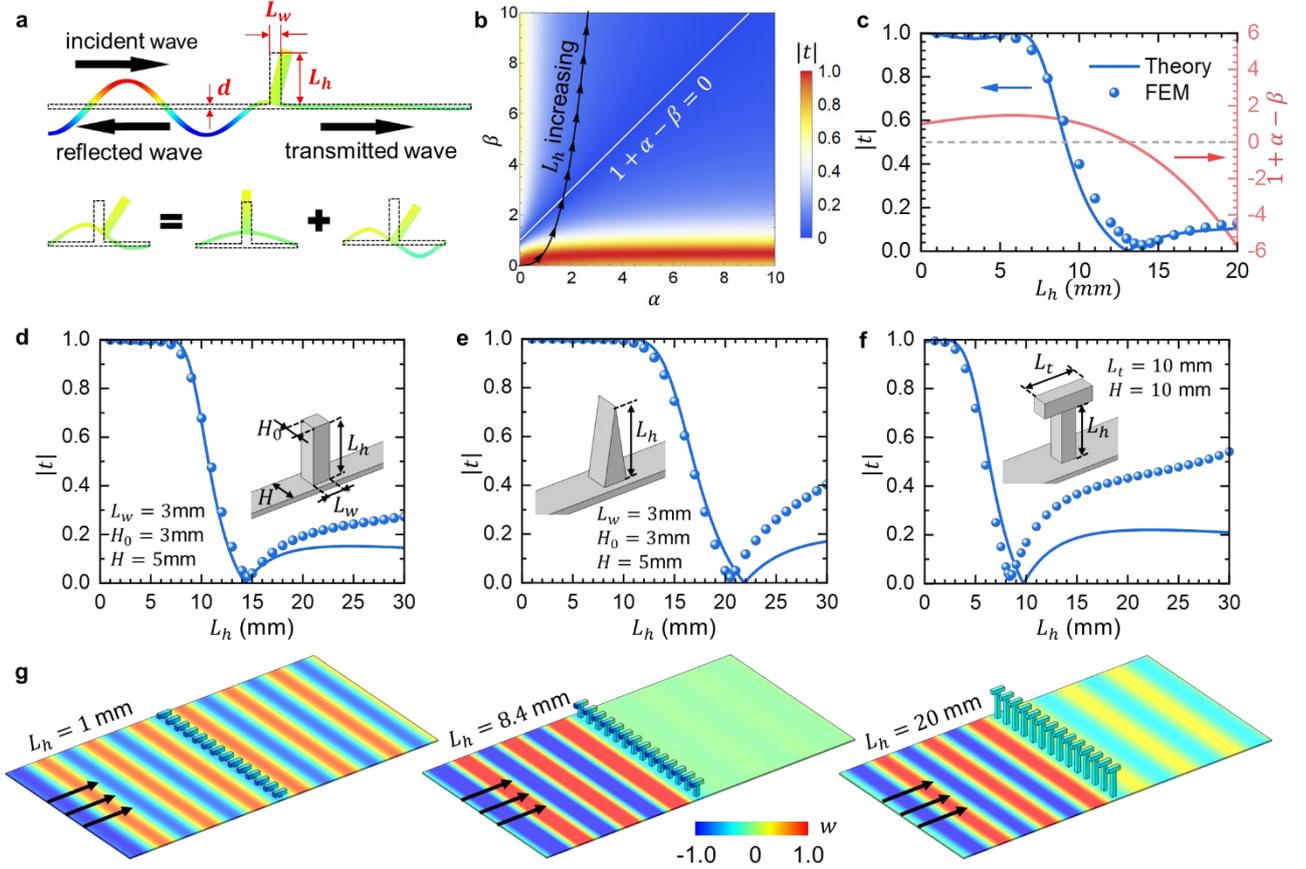

**Fig. 2 Transmission of flexural waves in a thin plate with a prominent obstacle. a** Schematic diagram of the 2D case. The motion of the obstacle can be divided into bending and twisting modes. **b** Density map of the transmission $|t|$ with respect to the relative dynamic stiffnesses $\alpha$ and $\beta$. $|t|$ is zero when $1 + \alpha - \beta = 0$ (the white line). The black curve with arrows corresponds to the variation of $\alpha$ and $\beta$ along with the increase of $L_1$ in **a**. **c** Transmission $|t|$ and the $1 + \alpha - \beta$ value as a function of $L_h$ for the 2D case. Curves and points correspond to the theoretical and numerical results, respectively. **d-f** Transmission $|t|$ as a function of $L_h$ for 3D cases with discrete rectangular, triangular and T-shape obstacles, respectively. **g** Numerical flexural wave fields with different T-shape obstacles. At $L_h = 8.4$ mm, the T-shape obstacles become meta-fences to totally reflect flexural waves.

To make it easier to assemble, the continuous "wall" can be discretized into 3D "fences", as illustrated in Fig. 2d. Moreover, we can change the shape of scatters at will, such as the triangular and the T-shape meta-fences in Fig. 2e and f, respectively. In all cases, both theoretical predictions and numerical results infer that properly designed scatterers can serve as meta-fences. In a more intuitive way, we further provide flexural wave fields at different heights $L_h$ in Fig. 2g. For a small value $L_h = 1.0$ mm, a high value of transmission $|t| = 0.996$ occurs, while a minimal transmission of $|t| = 0.029$ at $L_h = 8.4$ mm is observed as a meta-fence. With the further increase of $L_h$ to 20.0 mm, the transmission is degraded to $|t| = 0.432$. This trend clearly shows the characteristics of the proposed meta-fences: a blind oversized obstacle is not necessarily effective to stop the propagation of flexural waves, while a small but appropriate one can totally reflect it.

**Omnidirectionally wave-blocking with sparse arrangement**

For 3D cases with discrete meta-fence units, apart from the height $L_h$, the spacing of each discrete obstacle $H$ is also a key factor to block waves. As an example, we choose the horn-like shape in Fig. 3a as the basic structure of the unit cell, whose profile can be sketched by two arc-sine curves as

$$\begin{cases} x(s) = s \pm \dfrac{L_w}{2L_t}(s - L_t) \\ z(s) = \dfrac{L_h}{\pi}\arcsin\left(\dfrac{2s}{L_t} - 1\right) + \dfrac{L_h}{2} \end{cases} \quad (2)$$

where the argument $s \in [0, L_t]$, with the length $L_t = 20$ mm, the height $L_h = 20$ mm and the bottom-edge width $L_w = 10$ mm, respectively. Units with thickness $H_0 = 3.0$ mm are periodically arranged at a sparse interval of $H$. It is noteworthy that the proposed horn-like unit shows advantages over other shapes due to lighter weight and sparser layout compared with the regular rectangular unit to achieve omnidirectional reflection (Supplementary Notes 1 and 4). In Fig. 3b, we numerically calculate the transmission $|t|$ and $1 + \alpha - \beta$ as a function of $H$ (see Supplementary Note 2 for detailed procedures). As expected, the transmission approaches to zero ($|t| = 0.01$) exactly when $1 + \alpha - \beta = 0$ at $H = 32.0$ mm, which is more than 10 times of $H_0$. Nevertheless, a denser arrangement of units may cause a higher transmission compared with a smaller interval of $H$.

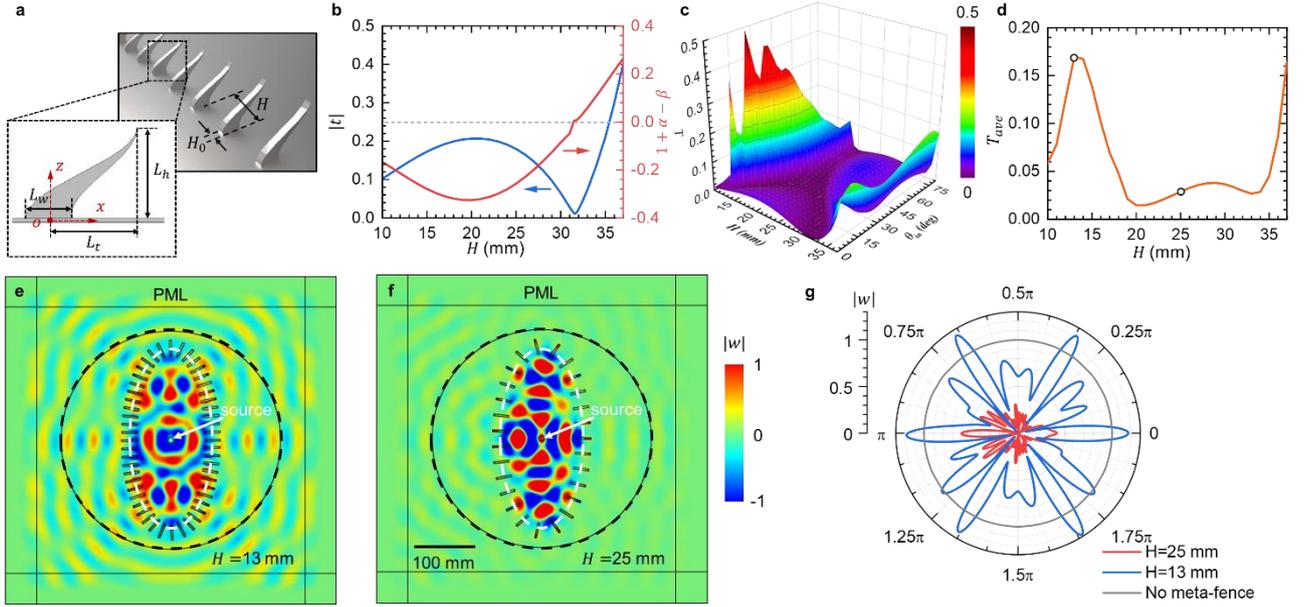

**Fig. 3 Design of a meta-fence to omnidirectionally isolate flexural waves in plates. a** Schematic of a meta-fence with discrete barriers. The horn-like units with thickness $H_0 = 3.0$ mm are periodically arranged at intervals of $H$. **b** Transmission $|t|$ and $1 + \alpha - \beta$ as a function of $H$. **c** Total transmissivity $T = |t|^2$ as a function of $H$ and $\theta_{in}$. **d** Averaged transmission $T_{ave}$ of different incident angles as a function of $H$. **e** Flexural wave field by placing a point source at the center of an elliptical meta-fence with semi-major axis 145 mm, semi-minor axis 61 mm and $H$ =13 mm, respectively. **f** Same as **e**, but with $H$ =25 mm. **g** Amplitude of the displacement $|w|$ out of the meta-fence (along the dashed black circle shown in **e** and **f**).

As a further step, to omnidirectionally isolate waves using the meta-fence, we numerically investigate the total transmissivity $T = |t|^2$ with different incident angles $\theta_{in}$. As plotted in Fig. 3c, the transmission for oblique incidence can be quite different from the normal incident case, but $T$ keeps a small value for $H \in [20, 30]$ mm regardless of the incident angle $\theta_{in}$. Additionally, the averaged transmissivity $T_{ave} = \frac{1}{n}\sum_{k=1}^{n} T_k$ is defined to characterize the overall performance of reflection in Fig. 3d. In fact, the total transmissivity $T$ is composed of the $0^{th}$ and $-1^{th}$ orders (Supplementary Note 3). $T_{ave}$ remains a stably small value, less than 0.03, when $H$ is in the region from 18.6 to 34.4 mm. From a stability point of view, the $H$ can be selected as 25 mm to design an omnidirectionally reflected meta-fence. It is noted that, similar to the normal incident case, a densely arranged structure is not the optimal wave isolator.

We then perform full wave simulations of the meta-fences by choosing two different intervals $H = 13$ and $25$ mm, respectively. As shown in Fig. 3e and f, the meta-fences are ensembled into an elliptical region with semi-major and semi-minor axes 145 mm and 61 mm, respectively. A point source is set to generate cylindrical flexural wave and to interact with the meta-fences at different incident angles. Few flexural waves can propagate through the meta-fences for $H = 25$ mm in Fig. 3f, while wave fields with distinct magnitudes can be observed for $H = 13$ mm in Fig. 3e, showing a good agreement with the results in Fig. 3d. The worse wave insulation performance of $H = 13$ mm attributes to higher transmission at oblique incident angles (Supplementary Note 5). In a quantitative way in Fig. 3g, we investigate the relative displacement amplitude $|w|$ along the dashed black circle shown in Fig. 3e and f. The mean values of $|w|$ are 0.683 and 0.201 for $H = 13$ mm and $H = 25$ mm, respectively. The magnitude $|w|$ for $H = 13$ mm even exceeds 1 at some angles due to multiple reflections. A better wave insulation performance for $H = 25$ mm than the case of $H = 13$ mm suggests that the period $H$ is a key index for our meta-fences.

**Design paradigm of omnidirectionally isolators and compact waveguides**

The omnidirectional wave reflection by our meta-fences may stimulate a variety of applications, for example, an efficient isolator of flexural waves as shown in Fig. 4a and b. By setting $H = 25$ mm, the meta-fence can be arranged into an arbitrary shape, which is chosen to be a quadrangle here. If we place a point source outside the isolator in Fig. 4a, waves almost bounce back and form ripples. The isolation effect can be quantitatively described by the amplitude of displacement on a straight line across the source. Before approaching the meta-fence ($x < -93$ mm), the average displacement is as large as 0.4. Then an abrupt drop in the displacement to around 0.1 occurs immediately upon reaching the meta-fence. Moreover, the value continuously drops to less than 0.01 in the shaded region ($x > 180$ mm) due to multiple barrier effects. The meta-fence isolator is expected to work well if we put the source inside it as shown in Fig. 4b. In this case, a strong and chaotic wave field is obtained and confined in the quadrangular region. The relative displacement colossally drops from 0.9 to 0.1 after passing the meta-fence interface. We further propose to design compact waveguides as depicted in Fig. 4c and d, where meta-fences are parallelly arranged or circularly arranged with width $L = 70$ mm. Beams propagate along the predetermined paths inside both waveguides, while negligible power escapes out of the waveguides. The relative amplitude at any point out of the waveguides is less than 0.1. Time domain simulation results for both isolators and waveguides are provided in Supplementary Movie 1. It should be emphasized that the present waveguides are quite compact (with $\sim\frac{\lambda}{2}$ thickness on each side), which shows great advantages than pervious approaches using defect modes of phononic crystals or topological interfaces[7-12].

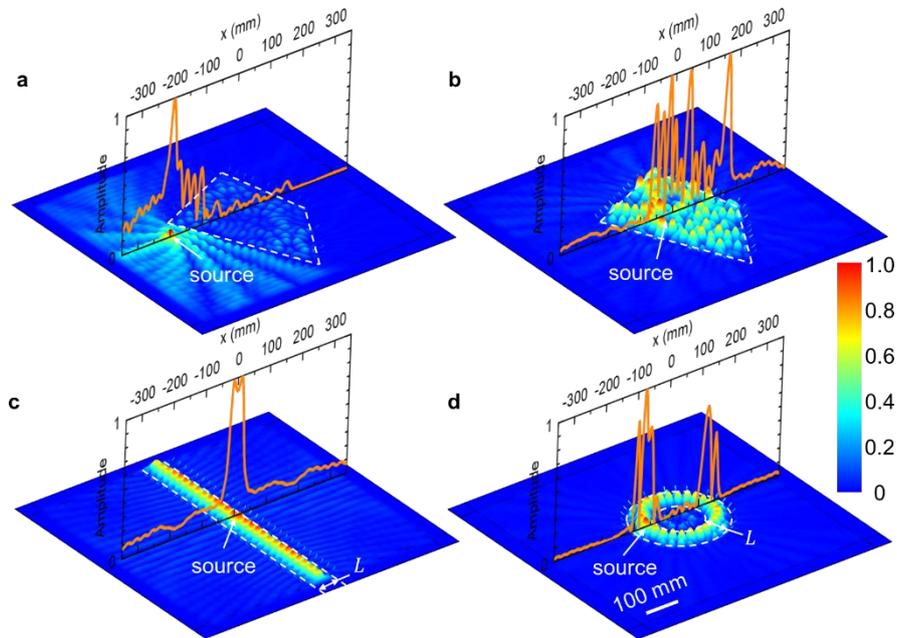

**Fig. 4 Simulated normalized amplitude of displacement with meta-fences at 5.5 kHz.** The amplitude on a typical line crossing the meta-fence (the orange curve) is examined in each subplot. A point source is placed outside **a** or inside **b** a quadrangular region surrounded by a meta-fence. **c** A short-line source propagates in a strip-like waveguide. **d** A point source propagates in an annular waveguide. In **c** and **d**, $L = 70$ mm.

**Experimental demonstration**

We further present the experimental verification of the above results, by arranging the meta-fence into a circular isolator (with radius 47.7 mm) through gluing 12 units onto a thin steel plate, as shown in Fig. 5a and Supplementary Movie 2. A point source at 5.5 kHz is stimulated to disperse a heap of glass beads (located to the left of the source), while the glass beads inside the meta-fence can stay in place to indicate excellent vibration isolation functionality. Our experimental results show that both the vibration isolation performance and the working frequency range are improved by a multi-layer configuration. In Fig. 5b, a small circular meta-fence (with radius 47.7 mm) is set to confine the point source excited by a piezoelectric patch, and a large one (with radius 151.2 mm) is used to prevent the leaked energy into the protected area. As expected, the flexural wave is well locked inside the small circle in shown in Fig. 5c. Reflection of the leaky wave by the large circular meta-fence is also observed, which enhances the isolation performance of the protected area. As a quantitative analysis in Fig. 5d, we choose two points ($B_1$ and $B_2$) at the same distance from the source, which are located in and out of the larger circular meta-fence, respectively. The value $\left|\frac{w_{B_2}}{w_{B_1}}\right| < 1$ in the frequency range of 4.2 to 5.8 kHz manifests the isolation performance of point $B_2$ by travelling through the large circle. Then the central point $A$ in the protected area is examined, where the red curve depicts the amplitude ratio of meta-fence over that without meta-fence. The value $\left|\frac{w}{w_0}\right|_A = 0.12$ at 5.5 kHz is always less than one from 3 to 7 kHz, implying that the isolator works well over a wide frequency range. These results highlight that the meta-fence shows great advantages than the metasurface approach[26], by strikingly increasing the working bandwidth and without any sacrifice of the bearing capacity of the plate.

In addition to isolators to stop waves and vibrations, the meta-fence can be assembled into a waveguide to indirect waves. As shown in Fig. 5f, we demonstrate a Z-shaped waveguide with $L = 70$ mm. Placing a point source in the upper right of the waveguide in Fig. 5e, one can see that the flexural wave is restricted to propagate along the pre-designed path over a long distance. Since the width $L$ is larger than the wavelength, the wave field in the waveguide looks distorted due to multiple reflections. By modulating the geometric parameters $H$ and $L$ as well as the working frequency, different guided modes are expected to be observed (Supplementary Note 6). The close-up view of displacement field in Fig. 5e is illustrated in Fig. 5g, showing an excellent wave-guiding performance and agreement with the numerical results.

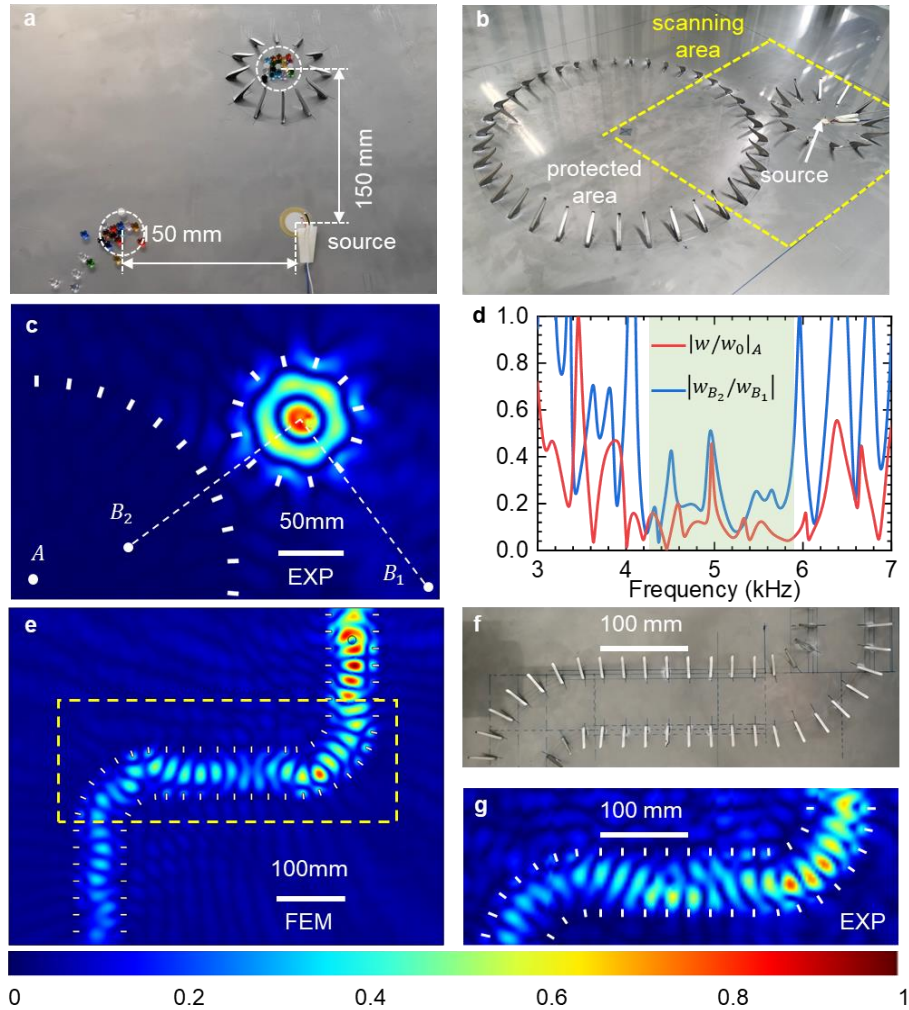

**Fig. 5 Experimental verification of isolators and waveguides made of meta-fences. a** Snapshot of a circular isolator working at 5.5 kHz. **b** Experimental setup of a two-circle isolator. **c** Measured amplitude of the displacement field at 5.5 kHz. **d** Tested isolating spectrum of the meta-fence. The red curve represents the amplitude ratio at point $A$ with meta-fence to that without meta-fence $|w/w_0|_A$. The blue curve refers to the amplitude ratio at two points ($B_1$ and $B_2$) at the same distance from the source $|w_{B_2}/w_{B_1}|$. **e** Simulated displacement field of a Z-shaped waveguide at 5.5 kHz. **f** Image of the waveguide. **g** Tested displacement field in the region framed by the red dashed box in **e**.

**Discussion**

We have demonstrated a high-efficiency meta-fence, consisting of a layer of identical tiny units glued on the plate, to totally block flexural waves using an appropriated-sized obstacle. Counter-intuitively, the meta-fence can omnidirectionally block waves by scatters with extremely small sizes and light weight. It is rationalized by the counter-balance effect among the incident wave, the restraining force, and the restraining moment of the obstacle. Simulations and experiments demonstrate that proposed meta-fences with horn-like units can be glued on the plate to form arbitrary-shaped isolators as well as waveguides.

Our meta-fence may exhibit distinct advantages to stimulate a number of potential applications. First, we can replace the unit cells of meta-fences by nonlinear[36] or local-resonating ones[37] to decrease the working frequency from kilohertz to tens of hertz for vibration control, which is of great importance in large mechanical and aerospace structures. Damping layers[28] can be also introduced into the meta-fence to enhance the vibration isolation performance. Second, the meta-fence is simpler and more flexible to form arbitrary-shaped waveguides without designing complex chambers and dense lattice units, which may enrich the toolbox of acoustic tweezers for drug delivery, cell engineering, and chemical reaction control[38,39]. Finally, the present scenario can also be extended to manipulate other types of waves, such as isolating Rayleigh waves in the ultrasonic frequency range for non-destructive testing and/or in the low frequency range for earthquake protection[40,41].

**Methods**

**Numerical simulations.** To simulate transmittances under normal and oblique incidence, we apply perfectly matched layers (PMLs) on both ends of beam-like models to absorbing reflections. The periodic boundary condition is used at the lateral boundaries to simplify calculations. The 304 stainless steel is set throughout the work with density $\rho = 7900 \ \frac{\text{kg}}{\text{m}^3}$, Young's modulus $E = 200$ GPa and Poisson's ratio $\nu = 0.3$. The thicknesses of all plates are set as 0.91 mm. For numerical wave fields of isolators and waveguides, PMLs are also used on outer boundaries of plates.

**Experimental setup.** In all experiments, horn-like units are fabricated by wire EDM and glued on a 1000×1000×0.91 mm$^3$ flat plate to form designed structures. An arbitrary waveform generator (Rigol DG4062) produces a 3-cycle tone burst $w(t) = A_0 \left[1 - \cos\left(\frac{2\pi f_c}{3}t\right)\right] \sin(2\pi f_c t)$, where the central frequency is $f_c = 7.5$ kHz. A power amplifier (ATA-2022H) amplifies the signal and transfers it to the piezoelectric patch bonded on the plate. To reduce reflected waves, we cover each edge of the large plate with a layer of blue-tack. The plate is placed on pyramidal sound-absorbing panels to isolate vibrations from environments and satisfy the free boundary condition. The wave profiles are measured by a laser vibrometer (Polytec NLV-2500) whose laser head is fixed on an electric moving platform enabling it to move in $xy$-plane and scan testing regions. The clearance of adjacent measuring points is 4 mm, corresponding to about 10 points per flexural wavelength at 5.5 kHz. An oscilloscope collects and transfers wave profiles from the vibrometer. Then, we can obtain the normalized amplitude distributions of measured fields (328×228 mm$^2$ for the isolator and 576×180 mm$^2$ for the waveguide) by transforming the measured wave profile from the time to the frequency domain.

## Acknowledgements
This work was supported by the NSFC under grant No. 11902239, No. 11902262, and the CNNC Science Fund for Talented Young Scholars.

## Author contributions
Y.L. and Y.Z. conceived the original idea. Y.L. and T.W. supervised the project. Y.Z., Y.L. and B.L. performed the theoretical calculation and numerical simulation. Y.Z. and G.S. carried out the experiments and analyzed the data. P. J., H.Z., J. Y. and Z.S. helped with the theoretical interpretation. Y.Z., B.L. Z.S. and Y.L. wrote the manuscript. All authors contributed to scientific discussions and reviewed the manuscript.

# Additional information

**Supplementary Information** accompanies this paper at [URL will be inserted by publisher] for Supplementary Figures 1-7, Supplementary Table 1, Supplementary Notes 1-6 and Supplementary Movies 1-2.

**Competing interests**：The authors declare no competing interests.

# References

1. Kushwaha, M. S., Halevi, P., Dobrzynski, L. & Djafari-Rouhani, B. Acoustic band structure of periodic elastic composites. *Phys. Rev. Lett.* **71**, 2022 (1993).
2. Vasseur, J. O. et al. Experimental and theoretical evidence for the existence of absolute acoustic band gaps in two-dimensional solid phononic crystals. *Phys. Rev. Lett.* **86**, 3012 (2001).
3. Liu, Z. Y. et al. Locally resonant sonic materials. *Science* **289**, 1734-1736 (2000).
4. Lu, M., Feng, L. & Chen, Y. Phononic crystals and acoustic metamaterials, *Mater. Today* **12**, 34-42 (2009).
5. Zhu, R., Liu, X. N., Hu, G. K., Sun, C. T., & Huang, G. L. Negative refraction of elastic waves at the deep-subwavelength scale in a single-phase metamaterial, *Nat. Commun.* **5**, 5510 (2014).
6. Ma, G. & Sheng, P. Acoustic metamaterials: from local resonances to broad horizons. *Sci. Adv.* **2**, e1501595 (2016).
7. Torres, M., Montero de Espinosa, F. R., García-Pablos, D. & García, N. Sonic band gaps in finite elastic media: surface states and localization phenomena in linear and point defects. *Phys. Rev. Lett.* **82**, 3054 (1999).
8. Wang, Z., Zhang, Q., Zhang, K. & Hu, G. Tunable digital metamaterial for broadband vibration isolation at low frequency. *Adv. Mater.* **28**, 9857-9861 (2016).
9. Wang, T. T. et al. Collective resonances of a chain of coupled phononic microresonators. *Phys. Rev. Appl.* **13**, 014022 (2020).
10. Süsstrunk, R. & Huber, S. D. Observation of phononic helical edge states in a mechanical topological insulator. *Science* **349**, 47-50 (2015).
11. Mousavi, S. H., Khanikaev, A. B. & Wang, Z. Topologically protected elastic waves in phononic

metamaterials. *Nat. Commun.* **6**, 8682 (2015).

12. Wang, P., Lu, L. & Bertoldi, K. Topological phononic crystals with one-way elastic edge waves. *Phys. Rev. Lett.* **115**, 104302 (2015).

13. Yu, S. et al. Elastic pseudospin transport for integratable topological phononic circuits. *Nat. Commun.* **9**, 3072 (2018).

14. Yan, M. et al. On-chip valley topological materials for elastic wave manipulation. *Nat. Mater.* **17**, 993-998 (2018).

15. Fan, H., Xia, B., Tong, L., Zheng, S. & Yu, D. Elastic higher-order topological insulator with topologically protected corner states. *Phys. Rev. Lett.* **122**, 204301 (2019).

16. Yu, N. et al. Light propagation with phase discontinuities: generalized laws of reflection and refraction. *Science* **334**, 333-337 (2011).

17. Fu, Y. et al. Reversal of transmission and reflection based on acoustic metagratings with integer parity design. *Nat. Commun.* **10**, 2326 (2019).

18. Xie, Y. et al. Wavefront modulation and subwavelength diffractive acoustics with an acoustic metasurfaces. *Nat. Commun.* **5**, 5553 (2014).

19. Cheng, Y. et al. Ultra-sparse metasurface for high reflection of low-frequency sound based on artificial Mie resonances. *Nat. Mater.* **14**, 1013-1019 (2015).

20. Popa, B., Zhai, Y. & Kwon, H. Broadband sound barriers with bianisotropic metasurfaces. *Nat. Commun.* **9**, 5299 (2018).

21. Su, G., Zhang, Y., Liu, Y. & Wang, T. Steering flexural waves by amplitude-shift elastic metasurfaces. *J. Appl. Mech.* **88**, 051011 (2021).

22. Zhu, H. & Semperlotti, F. Anomalous refraction of acoustic guided waves in solids with geometrically tapered metasurfaces. *Phys. Rev. Lett.* **117**, 034302 (2016).

23. Liu, Y. et al. Source illusion devices for flexural lamb waves using elastic metasurfaces. *Phys. Rev. Lett.* **119**, 034301(2017).

24. Lee, H., Lee, J. K., Seung, H. M. & Kim, Y. Y. Mass-stiffness substructuring of an elastic metasurface for full transmission beam steering. *J. Mech. Phys. Solids* **112**, 577-593 (2018).

25. Li, B. et al. Efficient asymmetric transmission of elastic waves in thin plates with lossless metasurfaces. *Phys. Rev. Appl.* **14**, 054029 (2020).

26. Zhu, H., Walsh, T. F. & Semperlotti, F. Total-internal-reflection elastic metasurfaces: design and application to structural vibration isolation. *Appl. Phys. Lett.* **113**, 221903 (2018).

27. Sua, X., Lu, Z. & Norris, A. N. Elastic metasurfaces for splitting SV- and P-waves in elastic solids.

*J. Appl. Phys.* **123**, 091701 (2018).

28. Cao, L. et al. Flexural wave absorption by lossy gradient elastic metasurfaces. *J. Mech. Phys. Solids* **143**, 104052 (2020).

29. Kim, M. S., Lee, W., Park, C. I. & Oh, J. H. Elastic wave energy entrapment for reflectionless metasurfaces. *Phys. Rev. Applied* **13**, 054036 (2020).

30. Ra'di, Y., Sounas, D. L. & Alù, A. Metagratings: beyond the limits of graded metasurfaces for wave front control. *Phys. Rev. Lett.* **119**, 067404 (2017).

31. Paniagua-Dominguez, R. et al. A metalens with a near-unity numerical aperture. *Nano Lett.* **18**, 2124-2132 (2018).

32. Chen, W. T., Zhu, A. Y. and Capasso, F. Flat optics with dispersion-engineered metasurfaces. *Nat. Rev. Mater.* **5**, 604-620 (2020).

33. Preumont, A. *Vibration Control of Active Structures: An Introduction*. (Springer, Netherlands, 2011).

34. Su, Z., Ye, L. & Lu, Y. Guided Lamb waves for identification of damage in composite structures: A review. *J. Sound Vib.* **295**, 753-780 (2006).

35. Mead, D. J. *Passive Vibration Control*. (John Wiley and Sons Inc, 1998).

36. Fang, X., Wen, J., Bonello, B., Yin, J. & Yu, D. Ultra-low and ultra-broad-band nonlinear acoustic metamaterials. *Nat. Commun.* **8**, 1288 (2017).

37. Morvaridi, M., Carta, G. & Brun, M. Platonic crystal with low-frequency locally-resonant spiral structures: wave trapping, transmission amplification, shielding and edge waves. *J. Mech. Phys. Solids* **121**, 496-516 (2018).

38. Li, J., Shen, C., Huang, T. J. & Cummer, S. A. Acoustic tweezer with complex boundary-free trapping and transport channel controlled by shadow waveguides. *Sci. Adv.* **7**, eabi5502 (2021).

39. Bourquin, Y., Wilson, R., Zhang, Y., Reboud, J. & Cooper, J. M. Phononic crystals for shaping fluids. *Adv. Mater.* **23**, 1458-1462 (2011).

40. Blitz, J., & Simpson, G. *Ultrasonic methods of non-destructive testing (Vol. 2)*. (Springer Science & Business Media., 1995).

41. Colquitt, D. J., Colombi, A., Craster, R. V., Roux, P., & Guenneau, S. R. L. Seismic metasurfaces: Sub-wavelength resonators and Rayleigh wave interaction. *J. Mech. Phys. Solids* **99**, 379-393 (2017).